\documentclass[11pt]{article}
\usepackage{moriond,epsfig}

\begin{document}
\vspace*{4cm}
\title{Planetary ephemerides and gravity tests in the solar system}

\author{A. Fienga}

\address{Institut UTINAM, 41 bis avenue de l'observatoire\\
Besan\c con, France}

\maketitle\abstracts{We review here the tests of fundamental physics based on the dynamics of solar system objects.
}

\section{Introduction}
\begin{figure}
\includegraphics[scale=0.4]{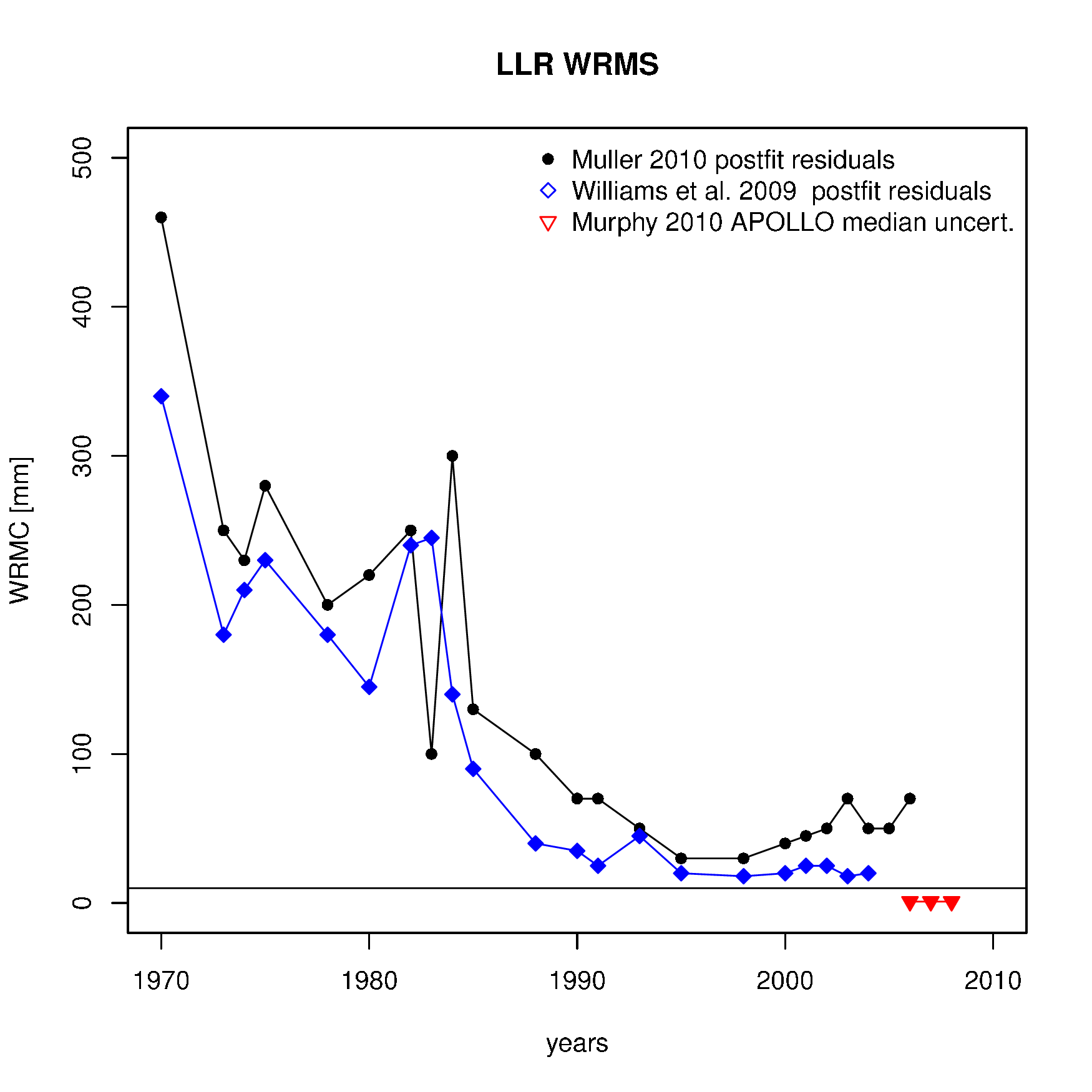}\includegraphics[scale=0.4]{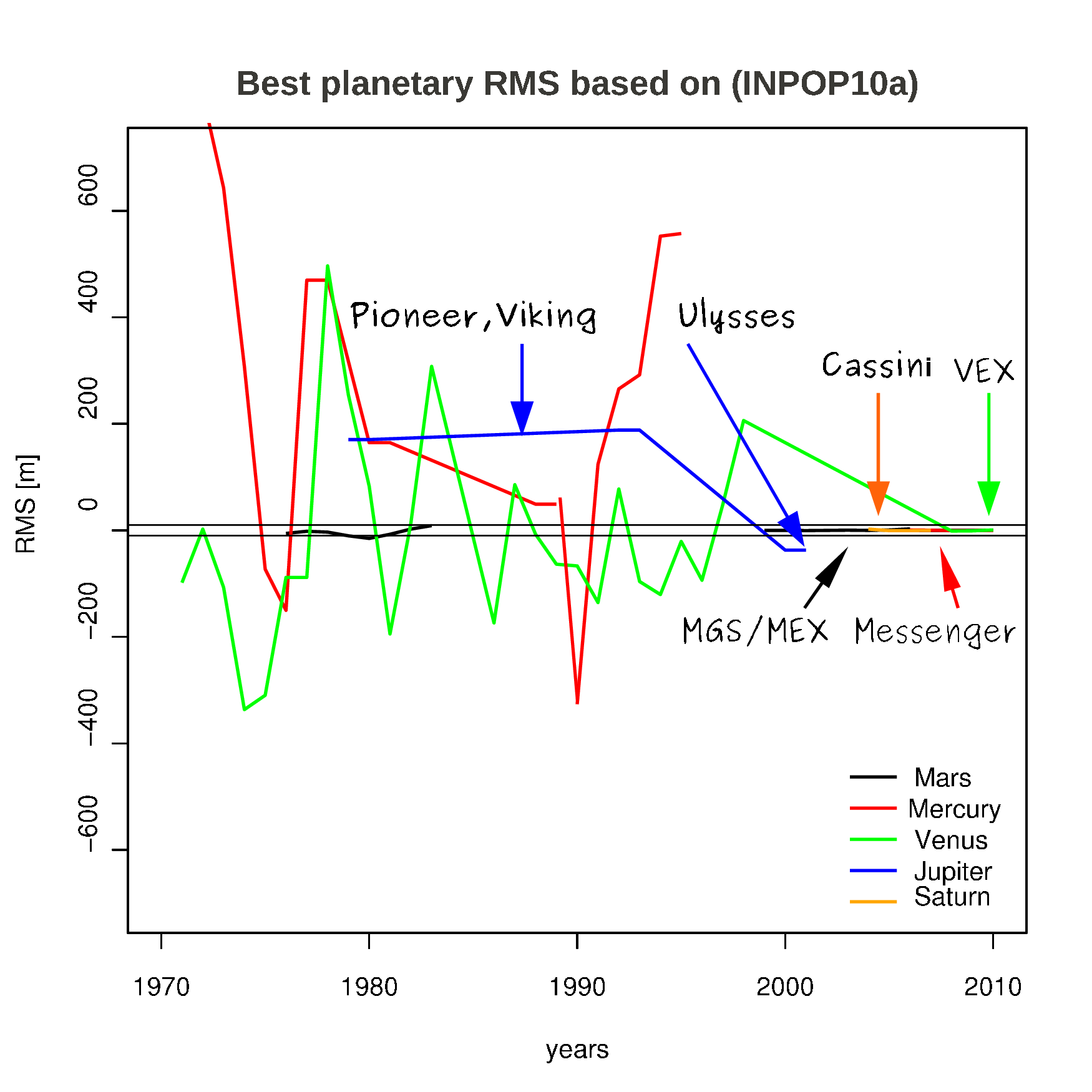}
\caption{Best fitted residuals for Moon and planets. The planetary residuals have been estimated with INPOP10a (\protect\cite{Fienga2011}) when the Moon residuals labelled (Muller 2010) have been estimated by \protect\cite{Muller2010} and those labelled (Williams et al. 2009) have been presented by \protect\cite{Murphy2010} as well as the APOLLO median uncertainties.}
\label{accuracy}
\end{figure}

The solar system has the most well known dynamics with a very ancient history of observations. GR (GR) was first tested at the astronomical scale since the dynamical impact of GR has then astronomical effects. The Eddington first observations of the light bending during the 1919 solar eclipse and the explanation by GR of the perihelia advance of Mercury in 1915 are the first stones of regular checks obtained at the astronomical scale with the Moon, the planets and the asteroids of the solar system. 
Since the nineties, planetary dynamics was drastically improved thanks to the numerous space missions orbiting planets and very accurately tracked by radio and VLBI observations.
Since the Apollo missions, the Moon is also intensively observed with even greater accuracy. 
The figure \ref{accuracy} pictures the situation.
For the Moon, the accuracy is at the centimeter level on a period of 35 years. For the planets, the geocentric distances are estimated at the meter level for Mars on a period of about 10 years due to MGS, MO and MEX tracking data. For Venus the improvement is also important but over a more limited period (2006-2010) thanks to the VEX mission. The improvement of the accuracy of Jupiter orbit stops with the latest available flyby data obtained with the Cassini mission in 2000. This latest gives very good constraints on the Saturn orbit which is now known with an accuracy of 20 meters over 3 years. 
GR is then confronted with very accurate observed positions of the Moon and the planets.
The solar system is then an ideal laboratory for testing gravity.

In the same time, theoretical developments ask to be tested in the solar system or forecast GR violations at the solar system scales. One can cite for example the violation of the equivalence principal by the standard models for unification of quantum physics and GR, the supplementary advances of planet perihelia and nodes expected by the MOND theories, the variation with distance of the PPN parameters $\beta$ and $\gamma$ induced by dark energy (\cite{2010AIPC.1241..690A}) or string theory (\cite{1996CQGra..13A..33D}, \cite{KR96}), variation of the gravitational constant $G$ induced by dark energy (\cite{2009PhRvD..79j4026S}) or scalar field theories (\cite{2003AnHP....4..347U}) as well as supplementary accelerations due to dark matter (\cite{2010AdSpR..45.1007A},\cite{1994ApJ...437..529N}), MOND theories (\cite{2011MNRAS.412.2530B}) or modified gravitational potentials (\cite{2010LRR....13....3D}, \cite{2010LRR....13....4T}).

\section{General overview}

\subsection{Tests based on direct spacecraft dynamics}

The accuracy of the spacecraft tracking data orbiting a planet, in interplanetary phase or during a flyby can reach up to few centimeters over several months. Such tracking is done using doppler shift observations or VLBI differential measurements which are very sensitive to the deflection of light. With such accuracy, the tracking data of space missions seem to be a good tool to test gravity. However, some factors as navigation unknowns (AMDs, solar panel calibrations), planet unknowns (potential, rotation...), effect of the
solar plasma, or the correlation with planetary ephemerides limit such gravity tests. Dedicated missions have then to be planed in order to overcome such difficulties. For example, the PPN $\gamma$ determination obtained by \cite{2003Natur.425..374B} was done during very specific periods of the Cassini mission, especially dedicated to such tests. 

The dynamics of the solar system planets and moons is less affected by non gravitational or unexpected accelerations and by technical unknowns and is constrained with also high accuracy.

\begin{table}
\caption{Results of gravity tests realized in the solar system. Columns 1 and 2 give the label of the tests and the studied objects. The Column 3 gives the obtained results with the mean value and 1 $\sigma$ least square standard deviations of the estimated parameter, except for the $\beta$, $\gamma$, $\dot{\varpi}_{\mathrm{sup}}$, and $\dot{\Omega}_{\mathrm{sup}}$ obtained with the planets. see the text for more details. The last column gives the alternative theories of gravity which can be constrained by these results. The values of $\beta$ given here were all obtained with a value of $\gamma$ given by \protect\cite{2003Natur.425..374B}. The ISL $\alpha$ obtained with LLR is for $\lambda = 4 \times 10^{8}$ km when the ISL $\alpha$ based on Mars data analysis is for $\lambda$ = $10^{10}$ km.}
\begin{tabular}{l c l l l}
\hline
Tests & Objects & Results & References & Theoretical impact\\ 
& & & & \\
\hline
EP $\eta$ $ \times 10^{4}$ & Moon-LLR & 4.4 $\pm$ 4.5& \cite{2009arXiv0901.0507083} & Standard model, \\
& & 6 $\pm$ 7& \cite{2008JGeod..82..133M} & string theory\\
$\left[ \Delta M /M \right] _{SEP} \times 10^{13}$ &  Moon-LLR & -2 $\pm$ 2 & \cite{2009arXiv0901.0507083} & \\
\hline
PPN $\gamma$ $ \times 10^{4}$ & Spacecraft & 0.21 $\pm$ 0.23 & \cite{2003Natur.425..374B} &Dark Energy,\\
with $\beta=1$& Planets  & 0.45 $\pm$ 0.75 &this paper & string theory\\
& Moon-LLR & 40 $\pm$ 50& \cite{2008JGeod..82..133M} &\\
PPN $\beta \times 10^{4}$ & Moon-LLR & 1.2 $\pm$ 1.1 & \cite{2009arXiv0901.0507083} & Dark Energy, \\
 & Moon-LLR  & 1.5 $\pm$ 1.8 & \cite{2008JGeod..82..133M} & string theory\\
 & Planets  & -0.41 $\pm$ 0.78 & this paper & \\
 & Planets  & 0.4 $\pm$ 2.4 & \cite{2011Icar..211..401K} & \\
\hline
$\dot{G}/G \times 10^{13}$ &  Moon-LLR & 2 $\pm$ 7 & \cite{2008JGeod..82..133M} & scalar-field theory\\
$[ y^{-1} ]$ &  Planets & 0.1 $\pm$ 1.6  & \cite{2011Icar..211..401K} &  \\
&  Planets & -0.6 $\pm$ 0.4 & \cite{Pitjeva2010} & \\
$\alpha_{1} \times 10^{5} $ &  Moon-LLR & -7 $\pm$ 9 & \cite{2008JGeod..82..133M} & scalar-field theory\\
$\alpha_{2} \times 10^{5}$ &  Moon-LLR & 2 $\pm$ 2 & \cite{2008JGeod..82..133M} &  \\
ISL $\alpha$ &  Moon-LLR & $10^{-10}$ & \cite{2008JGeod..82..133M} & Standard model\\
& Planets  &  $ 10^{-10}$ & \cite{2011Icar..211..401K} & \\
\hline
$\dot{\Omega}_{\mathrm{de Sitter}}$ &  Moon-LLR &  6 $\pm$ 10  & \cite{2008JGeod..82..133M} & MOND, Dark matter\\
$[$mas.cy$^{-1}]$ & & & & \\
$\dot{\varpi}_{\mathrm{sup}},\dot{\Omega}_{\mathrm{sup}}$ & Planets  &  40 $\rightarrow$ 0.1  & this paper & \\ 
$[$mas.cy$^{-1}]$ & & & & \\
$ a_{supp}$ & Moon-LLR & $10^{-16}$ & \cite{2008JGeod..82..133M} & Dark Matter density\\
$[$m.s$^{-2}]$& Planets  &  $ 10^{-14}$ & this paper, \cite{} & Pioneer anomaly\\
\hline
\end{tabular}
\label{tests}
\end{table}

\subsection{LLR tests}

With LLR observations, positions and velocities of the Moon are known with an accuracy from 10 to 1 centimeter over 35 years.
With the APOLLO project (\cite{2008PASP..120...20M}), new developments in observational techniques improve this accuracy with an observational error of about 1 millimeter. 
With such accuracy, \cite{2010LRR....13....7M} plans improvements of at least one order of magnitude in the test of the equivalence principle, the determination of the PPN parameter $\beta$ and of the test of  inverse square law. 
The table \ref{tests} gathers the main tests of gravity done using LLR observations as well as planetary ephemerides and spacecraft tracking. The LLR analysis is clearly one of the main source of information for gravity. It produces tests with the best available accuracy for the equivalence principal, the prefered-frame tests and the detection of possible supplementary accelerations induced by dark matter.
Present limitations in the modeling of the lunar interior, the Earth rotation as well as the planetary ephemerides induce differences between the teams analyzing the LLR observations (\cite{2008JGeod..82..133M}, \cite{2009arXiv0901.0507083}) of several centimeters where the supposed accuracy of the APOLLO observations is about 1 millimeter. These discrepancies are  obvious on figure \ref{accuracy}. An intensive work of comparisons and improvement of the Moon dynamics is now in progress.
 
\subsection{Tests based on planetary ephemerides}

As already discussed in section 1, the tracking observations of spacecrafts orbiting or flying by planets became very accurate in the nineties.
The Mars data are crucial for the adjustment of the planetary ephemerides due to their high accuracy (about few meters on the Earth-Mars distances), their number (about 30 \% of the complete data set) and their long time span (from 1998 to 2010). This makes the planetary ephemerides very depend on this type of data and also very sensitive to the Mars orbit modeling and to the perturbations of the asteroids.
On table \ref{tests} are found the tests done with planetary ephemerides. The dynamics of the planets give very competitive estimations of $\beta$ and $\gamma$ PPN parameters as well as variations with time of the gravitational constant. Supplementary acceptable advances in the nodes and perihelia of planets are also constrained with the observations used in the adjustment of the planetary ephemerides. Tests of possible Pioneer-like accelerations on outer planet orbits have also been tested with high accuracy.
The limitations of the gravity tests obtained with planetary ephemerides are mainly linked with the overweight of the Mars datasets and with the perturbations of the asteroids. As demonstrated in \cite{2008JGeod..82..133M}, there is a strong correlation on the geocentric Mars distances between the sun oblatness, the PPN parameter $\beta$ and the mass of the asteroid ring used to average the perturbations of small asteroids on planet orbits. The decorrelation  between the PPN parameters, the sun oblateness and the asteroid perturbations is then better obtained when the global adjustement of all the planet orbits is done simultaneously.
Furthermore, the decorrelation between the $\beta$ and $\gamma$ parameters are only possible if the two following equations are solved together in the fitting procedure.


where $\Delta\dot\varpi$ is the advance of perihelia of the perturbed planet induced by GR (first term) and the oblateness of the sun ${J_2}$ (second term with $R^2_{\odot}$, the sun radius). In the first equation, $a$,$e$ are the semi-major axis and the eccentricity of the perturded planet and $GM_{\odot}$ is the mass of the sun. In the  second equation, $\Delta t$ is the Shapiro delay, the supplementary delay induced by the deflection of the light path by the sun. 
\section{Results and discussions}

\subsection{Equivalence principal}

A detailed description of the method used to test the equivalence principal in using LLR observations is given in \cite{2009arXiv0901.0507083}. The test is an estimation of the gravitational mass to inertial mass ratio of the Moon and of the Earth. In GR, this ratio is equal to 1. However in PPN formalism, it can be formulated as
$$
M_{G} / M_{I} = 1 + \eta [ (\frac{U}{Mc^{2}})]
$$
where $U$ is the body's gravitational self-energy, $M$ is the mass of the body, $c$ the speed of light and $\eta$ a PPN parameter equal to 0 in GR. In the equation of the geocentric motion of the Moon, \cite{2009arXiv0901.0507083} have introduced the differences in accelerations induced by $M_{G} / M_{I} \neq 1$ for the Moon and of the Earth. By comparisons to the LLR observations, it becomes possible to estimate the acceptable $M_{G} / M_{I}$ ratio by direct least square fit of the numerically integrated acceleration or of the analytical estimations based on \cite{1968PhRv..170.1186N}. The results presented in table \ref{tests} are the one obtained by \cite{2009arXiv0901.0507083} combined with the laboratory estimations of the weak equivalence principal obtained by \cite{2001CQGra..18.2397A}.

As the PPN parameter $\eta$ is linked to the $\beta$ and the $\gamma$ by $\eta = 4 \beta - \gamma -3$,
for a given value of $\gamma$, values for the $\beta$ parameters can be deduced. 
Same methods have been applied by \cite{2008JGeod..82..133M} with about the same results. Values obtained by \cite{2009arXiv0901.0507083} and \cite{2008JGeod..82..133M} are given in table \ref{tests}.

\subsection{PPN parameter $\beta$ and $\gamma$}

\begin{figure}
\begin{center}
\includegraphics[width=8cm]{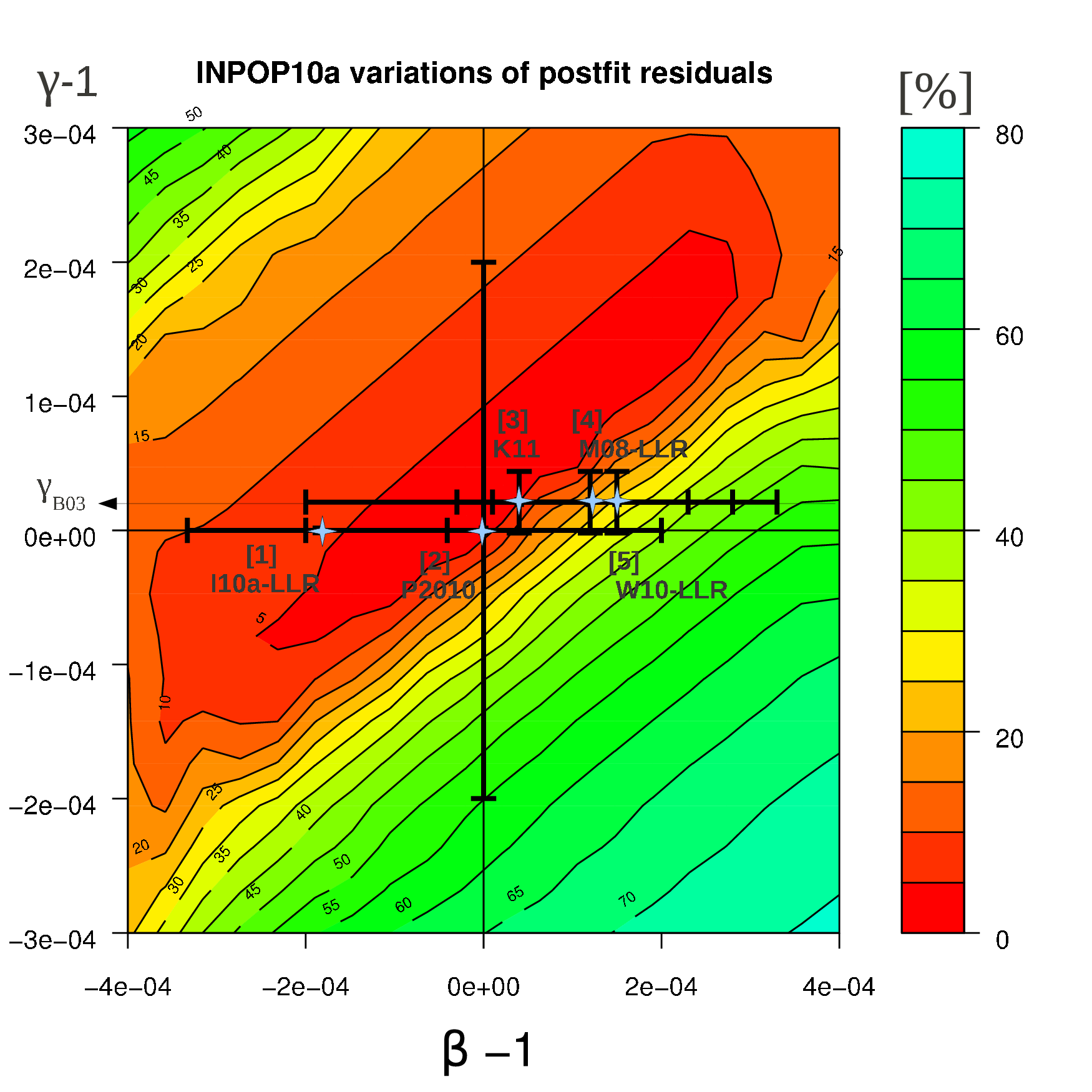}\includegraphics[width=8cm]{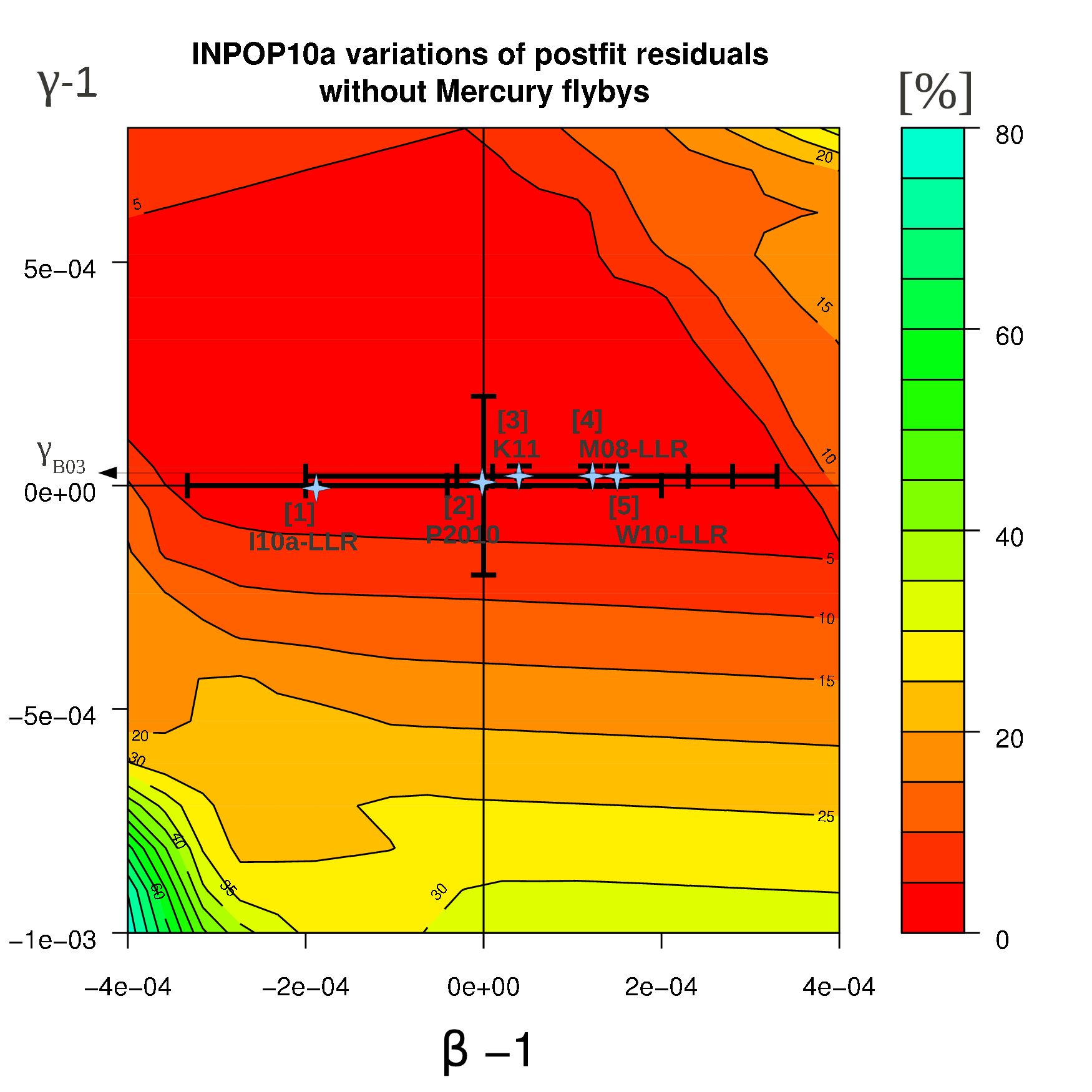}
\end{center}
\caption{Variations of postfit residuals obtained for different values of PPN $\beta$ (x-axis) and $\gamma$ (y-axis). [1] stands for a PPN $\beta$ value obtained by (\protect\cite{Manche2010}) using LLR observations with $\gamma=0$, [2] stands for \protect\cite{2009IAU...261.0603P} by a global fit of EPM planetary ephemerides. K11 stands for \protect\cite{2011Icar..211..401K} determinations based mainly on Mars data analysis. M08 for \protect\cite{2008JGeod..82..133M} and W09 for \protect\cite{2009arXiv0901.0507083} give values deduced from LLR for a fixed value of $\gamma$, B03 stands for \protect\cite{2003Natur.425..374B} determination of $\gamma$ by solar conjunction during the Cassini mission.}
\label{betagamma}
\end{figure}

Since \cite{Fienga2008} and \cite{2010IAUS..261..159F}, estimations of the PPN parameters are done with INPOP on a regular basis as well as estimations of acceptable supplementary advances of perihelia and nodes. A specific method presented in \cite{2010IAUS..261..159F} is used for these estimations.
In order to overcome the correlation problems, we built up several planetary ephemerides for different values of the PPN parameters ($\beta$, $\gamma$) with a simultaneous fit of initial conditions of planets, mass of the Sun and asteroid densities.
On figure \ref{betagamma} are plotted the variations of postfit residuals induced by the modification of the corresponding $\beta$ and $\gamma$ parameters. The left hand side plot gives the variations of postfit
 residuals including Mercury flyby normal point when the right hand side plot gives the variations of residuals without the Mercury observations.
The different levels of colors indicate the percentage of variations of the postfit residuals compared to those obtained with INPOP10a. By projecting the 5\% area on the $\beta$-axis (or the $\gamma$-axis), one can deduced the corresponding $\beta$ (or $\gamma$) interval given in table \ref{tests} in which the residuals are modified by less than 5\%.
In looking at the two figures, one can see that the use of the Mercury flyby data give smallest intervals of possible $\beta$,$\gamma$. This is consistent with the fact that the Mercury observations are far more sensitive to gravity modifications than other data (see table 1 in \cite{2010IAUS..261..159F} ).

\subsection{Frame-dragging and prefered frame tests}

Based on LLR observations, \cite{2008JGeod..82..133M} estimate a supplementary advance in the node of the Moon induced in GR by the motion of the Moon in the gravitational field of the Earth. The effect is called the de Sitter effect and the results are presented in table \ref{tests}.
Prefered-frame coefficients $\alpha_{1}$ and $\alpha_{2}$ have also been estimated by the same authors. Results are presented in table \ref{tests}.

\subsection{Supplementary accelerations}

\begin{figure}
\begin{center}
\includegraphics[scale=0.6]{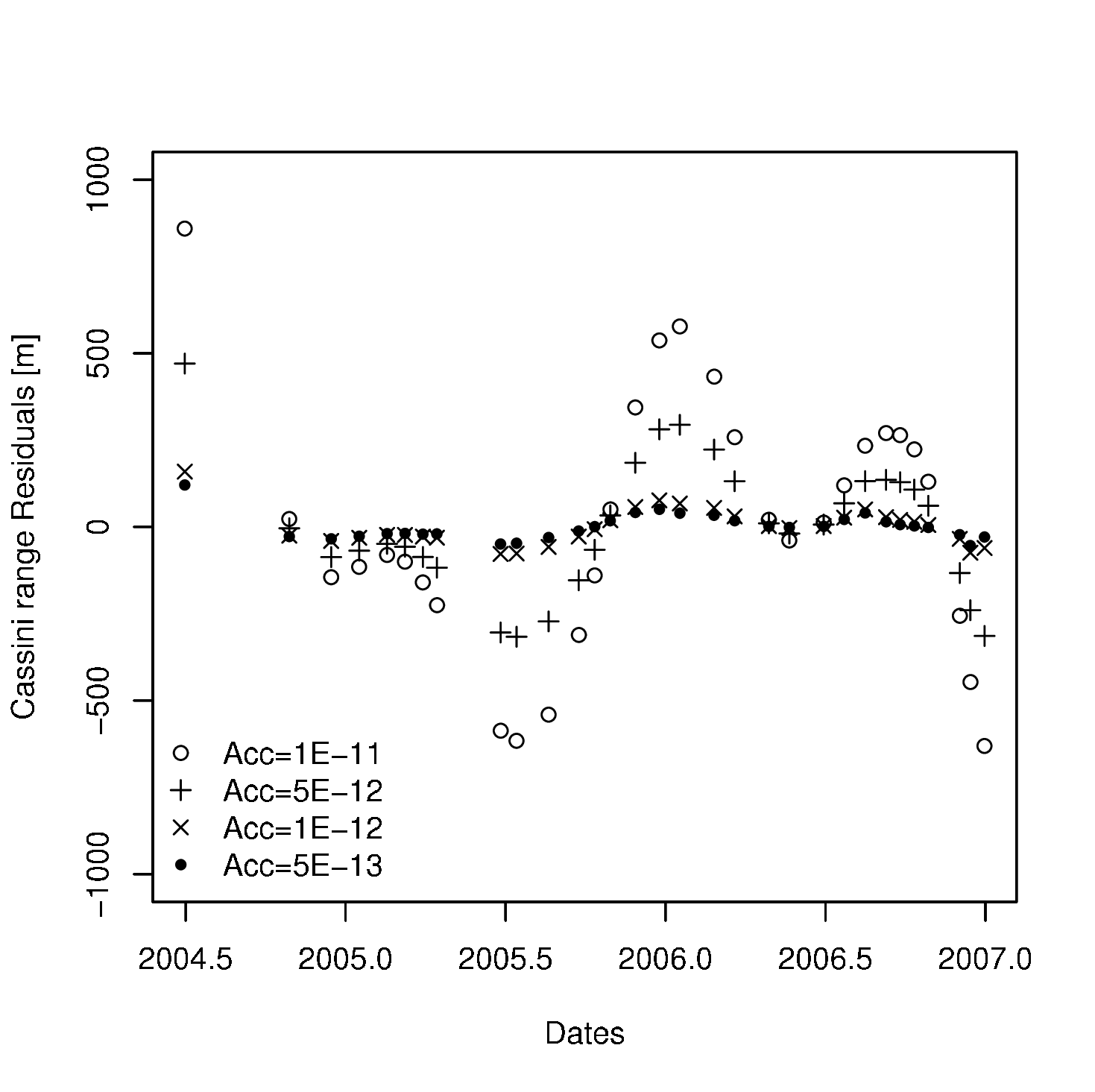} 
\caption{Postfit residuals in meters obtained by comparison between Cassini tracking data and several ephemerides with added Pioneer-like accelerations. The characteristics of the accelerations are given in m.s$^{-2}$}
\end{center}
\label{figpioneer}
\end{figure}

Supplementary accelerations can be induced by dark matter and dark energy surrounding the solar system or inside the solar system. Alternative descriptions of gravity can also induced modifications in the gravitational potential and then supplementary acceleration in the motion of solar system bodies and spacecraft.
Possible tests have been made in introducing either constant accelerations in one specific direction (Pioneer-like anomaly with outer planets, dark matter with the Earth-Moon orbit) either accelerations induced by f(r) gravity or exponential potentials (ISL).
\cite{2008JGeod..82..133M} and \cite{2011Icar..211..401K} have constrained the ISL potential for the geocentric Moon ($\lambda$ = $4 \times 10^{8}$ km) and Mars ($\lambda$ = $10^{10}$ km). 
Some other estimations should be investigate for $\lambda > 10^{12}$ km. Results are given in table \ref{tests}.
For the Pioneer-like accelerations, the figure \ref{figpioneer} gives the postfit residuals obtained by comparisons between the observed Earth-Saturn distances deduced from the Cassini tracking data and planetary ephemerides integrated with supplementary constant accelerations and fitted over the INPOP10a data sample. These accelerations are similar in direction to the Pioneer accelerations but their modules vary from 10$^{-11}$ to 5.10$^{-13}$ m.s$^{-2}$. As one can see on figure \ref{figpioneer} only accelerations smaller than 5.10$^{-13}$ m.s$^{-2}$ are acceptable compared to the present accuracy of the Cassini observations. This result was confirmed by \cite{2010IAUS..261..155F}.

\subsection{Supplementary advances of perihelia and nodes}
  
  New theoretical models (\cite{2011MNRAS.412.2530B}, \cite{2008PhRvD..77h3005F}) forecast supplementary advances in the orbits of the solar system objects.
  Supplementary advances in the perihelia and nodes of the planets (from Mercury to Saturn)  have also been tested in INPOP and are presented on table \ref{paramfitd}.
All these results are  based on the method presented in \cite{2010IAUS..261..159F}. 
With INPOP10a \cite{Fienga2011} no supplementary advances in node or perihelia is detected when with INPOP08 \cite{2010IAUS..261..159F} supplementary advance in the Saturn perihelia was possible. Such variations can be explained by the improvement of INPOP10a outer planet orbits compared to INPOP08.

\begin{table}
\caption{Values of parameters obtained in the fit of INPOP08 \protect\cite{2010IAUS..261..159F} and INPOP10a \protect\cite{Fienga2011} to observations. The supplementary advances of perihelia and nodes are estimated in INPOP10a and INPOP08 as the interval in which the differences of postfit residuals from INPOP10a are below 5\%. P09 stands for (\protect\cite{2009IAU...261.0603P}) and P10 for (\protect\cite{Pitjeva2010}).}
\begin{center}
\begin{tabular}{l c c c c}
\hline
  &INPOP08 & INPOP10a & P09& P10\\
\hline
$\dot{\varpi}_{\mathrm{sup}}$ &  & & &\\
mas.cy$^{-1}$ &  & & &\\
Mercury & -10 $\pm$ 30 & 0.4 $\pm$ 0.6 & -3.6 $\pm$ 5 & -4 $\pm$  5\\
Venus & -4 $\pm$ 6& 0.2 $\pm$ 1.5 & -0.4 $\pm$ 0.5 & \\
EMB & 0.0 $\pm$ 0.2 & -0.2 $\pm$ 0.9 & -0.2 $\pm$ 0.4 & \\
Mars & 0.4 $\pm$ 0.6 & -0.04 $\pm$ 0.15 & 0.1 $\pm$ 0.5 & \\
Jupiter & 142 $\pm$ 156 & -41 $\pm$ 42 & & \\
Saturn & -10 $\pm$  8 & 0.15$\pm$ 0.65 & -6 $\pm$  2 & -10 $\pm$  15\\ 
$\dot{\Omega}_{\mathrm{sup}}$  & & & &\\
mas.cy$^{-1}$&  & & &\\
Mercury & &  1.4 $\pm$ 1.8 & &\\
Venus & 200 $\pm$ 100 & 0.2 $\pm$ 1.5 & &\\
EMB & 0.0 $\pm$ 10.0 & 0.0 $\pm$ 0.9  & &\\
Mars & 0.0 $\pm$ 2 & -0.05 $\pm$ 0.13& &\\
Jupiter & -200 $\pm$ 100 & -40 $\pm$ 42 & &\\
Saturn & -200 $\pm$  100 & -0.1 $\pm$ 0.4  & & \\
\hline
\end{tabular}
\end{center}
\footnotetext[1]{fixed}
\label{paramfitd}
\end{table}

\section{Conclusions}

With the present gravity tests done in the solar system, GR is confirmed at the 10$^{-4}$ accuracy for PPN parameter $\beta$ and the equivalence principal and at 10$^{-5}$ for PPN $\gamma$. No supplementary advances of perihelia and nodes are detected at the present accuracy of the planetary ephemerides. Variations of the gravitational constant are not expected above 10$^{-13}$ yr$^{-1}$ and stringent limits for the ISL tests are given for the inner solar system. 
Messenger tracking data would bring important informations for PPN parameter determinations and ISL tests should be done in the outer solar system.

\section{REFERENCES}
%
%
%
%
%

\small{
\bibliography{biblio_hdr.bib}{}
}

\end{document}